\documentclass[fleqn,10pt]{wlscirep}
\usepackage[utf8]{inputenc}
\usepackage[T1]{fontenc}
\usepackage{bm}
\usepackage[cal=cm]{mathalfa}
\usepackage{tabularx}

\title{Noisy active matter}

\author[ ]{Atanu Chatterjee}
\author[ ]{Tuhin Chakrabortty}
\author[*]{Saad Bhamla}
\affil[ ]{Georgia Institute of Technology, School of Chemical and Biomolecular Engineering, Atlanta, United States}
\affil[*]{e-mail: saadb@chbe.gatech.edu}

\begin{abstract}
Noise threads every scale of the natural world. Once dismissed as mere background hiss, it is now recognized as both a currency of information and a source of order in systems driven far from equilibrium. From nanometer-scale motor proteins to meter-scale bird flocks, active collectives harness noise to break symmetry, explore decision landscapes, and poise themselves at the cusp where sensitivity and robustness coexist. We review the physics that underpins this paradox: how energy-consuming feedback rectifies stochastic fluctuations, how multiplicative noise seeds patterns and state transitions, and how living ensembles average the residual errors. Bridging single-molecule calorimetry, critical flocking, and robophysical swarms, we propose a unified view in which noise is not background blur but a tunable resource for adaptation and emergent order in biology and engineered active matter.
\end{abstract}

\begin{document}

\flushbottom
\maketitle
\thispagestyle{empty}

\section*{Introduction}
The story of noise is the story of modern science. From the background chatter that muddles conversation to the thermal jiggle of molecules that seeds genetic mutations, noise permeates everyday life and the living alike. Once a purely sonic notion, it soon became a cornerstone of fluctuation and information theory. This review traces that transformation (\textbf{Box 1}), showing how systems far from equilibrium treat noise as a resource that builds structure and function. Nowhere is this interpretation more vivid than in \textit{active matter}. 

From motor proteins to bird flocks, active matter comprises self-propelled units that expend energy to interact. Because power is supplied locally, the collective never equilibrates; instead it swarms, patterns, and flips between distinct collective states. Weaving the thread of noise through this fabric of active matter, we pose a deceptively simple question: if noise permeates every scale of living matter, can it reveal a unifying perspective on life? 

We argue that it can. First, across scales, we identify the conditions under which fluctuations become constructive, seeding symmetry-breaking, pattern formation, and collective decisions. Second, we quantify the energetic price life pays for harnessing randomness, using stochastic thermodynamics as the ledger. Tracing noise from molecular machines to organismal swarms, we expose design principles that unite a strikingly broad class of active systems. 

At the molecular scale, thermal agitation incessantly buffets nanometer-sized components. Rather than quelling these random kicks, motor proteins, gene circuits, cellular collectives, and self-assembling polymers act as Maxwell's demons: they sense each fluctuation, spend an ATP phosphate coin to bias reaction pathways, and ratchet the motion into directed work. Noise becomes information, and information becomes work, binding thermodynamic cost to cellular function at every step.

Zooming out a few orders of magnitude, the solitary bookkeeper morphs into an orchestrator: a roaming sensor-actuator loop that listens to sensory, social, and environmental noise and weaves it into collective choice by spending behavioral coins. Each tap of an antenna or a drop of a pheromone inscribes a small energetic cost in the collective memory, and as many such coins accumulate in this distributed ledger, noisy local interactions coalesce into a shared consensus.

Swarms of coin-sized robots now close the loop between biology and active matter theory by serving as a tractable ledger on which every energetic coin and transaction can be audited in real time. These programmable swarms are stand-ins for cellular collectives and organismal groups: they realize the same interaction primitives yet let us tune noise, dissipation, delays, and topology while auditing information flow in real time, turning the swarm into a synthetic echo of Maeterlinck's \textit{spirit of the hive}~\cite{maeterlinck1901life}.

In what follows, we treat noise as a resource across scales: from molecules to cells, where dissipation and feedback rectify fluctuations to sustain function, and ultimately survival in noisy, fluctuating environments; to organisms and groups, where noise-driven multistability and heightened susceptibility yield collective decisions essential for survival in uncertain landscapes, as we close with robophysical tests and design rules that adjudicate mechanisms and design rules, outlining the next frontier.

\section*{Noise at the molecular scale}
Living cells rely on a steady influx of free energy, often from ATP hydrolysis, to break detailed balance and convert random fluctuations into directed processes such as  protein synthesis~\cite{allan2009evolutionary},  signal transduction~\cite{qian2005nonequilibrium}, and molecular transport~\cite{Alberts2002-vo}. Already in 1944, Schr\"odinger framed this free-energy inflow as \textit{negentropy}, the improbable currency that living matter must keep spending to stave off thermal equilibrium~\cite{schrodinger1992life}. In practice, that free energy powers a hierarchy of molecular mechanisms that regulate noise: gene circuits parse biochemical shot noise to make probabilistic growth, differentiation, or survival decisions~\cite{tsimring2014noise}, motor proteins rectify thermal agitation into mechanical work, and cytoskeletal networks harvest local fluctuations to self-organize into adaptable architectures.

    \begin{center}
    \textbf{Box 1 \textbar\ Echoes of noise: a journey through physics~\cite{yeang2023transforming}}
    \fbox{
    \begingroup
    \footnotesize            
    \begin{minipage}{0.95\textwidth}
    \paragraph{\textit{From acoustic interference to thermal jitter}}\mbox{}\\[1pt]
    During the Industrial Revolution, advances in wave theory and mathematical analysis reshaped our understanding of sound, light, and heat~\cite{musson1989science}. Fourier's 1822 treatise demonstrated that any complex waveform can be decomposed into simple sinusoidal components, a pivotal breakthrough in the study of heat transfer and vibration~\cite{fourier1888theorie}. Doppler's 1842 work on frequency shifts further elucidated how motion alters the waves we observe~\cite{doppler1903ueber}, while Kirchhoff subsequently generalized the governing equations for wave propagation in arbitrary media~\cite{kirchhoff1857ueber}. These theoretical advancements paved the way for transformative inventions such as Bell's telephone (1876) and Edison's phonograph (1877)~\cite{bell1908bell,edison1888perfected}. Yet that very success laid bare a stubborn limitation: background noise distorted signals and muffled sound, turning a once-ignored nuisance into a pressing scientific problem. 
    
    \paragraph{\textit{Brownian ballet and the statistical view of nature}}\mbox{}\\[1pt]
    While engineers struggled to tame macroscopic noise in telegraph lines, a very different riddle surfaced under the lens of the Scottish botanist Robert Brown. In 1827, Brown reported that pollen grains suspended in still water moved with incessant, erratic jitters despite the absence of obvious external forces~\cite{brown1828xxvii}. His observation drew immediate speculation from luminaries such as Faraday, Brewster, and Darwin, who proposed capillary flows, light pressure, or electrical forces. Yet, by the 1870s, none of these classical explanations held up to experiment. This relentless motion, soon christened \textit{Brownian}, puzzled scientists for decades, hinting that noise could be intrinsic to matter. 
    
    The breakthrough came in the early twentieth century. Working independently, Einstein (1905) and Smoluchowski (1906) showed that Brown's wandering grains could be understood as the visible consequence of countless molecular collisions in the surrounding fluid~\cite{einstein1905motion,smoluchowski1906zusammenfassende}. Building on the statistical foundations laid several decades earlier by Gibbs and Boltzmann~\cite{gibbs1902elementary,sharp2015translation}, Einstein, in one of his \textit{Annus mirabilis} papers, modeled these random kicks as a stationary stochastic process and derived the diffusion law linking microscopic thermal fluctuations to macroscopic particle displacement. His analysis provided the first quantitative measure of Avogadro's number and, more broadly, converted Brownian noise into quantitative proof of molecular nature of matter~\cite{perrin1910mouvement}. By idealizing the forces as delta-correlated white noise, Einstein also laid the conceptual groundwork for modern stochastic calculus and the study of fluctuation-driven phenomena across physics, chemistry, and biology. 
    
    \paragraph{\textit{From thermal fluctuation to information entropy}}\mbox{}\\[1pt]
    By the 1920s, attention had shifted from Brownian motion in fluids to electrical circuits and precision instruments. Working with ultra-sensitive galvanometers, Zernike (1927) observed fluctuations that could not be attributed to any mechanical disturbances and correctly identified them as resulting from the thermal agitation of electrons~\cite{zernike1932brownsche}. Within a year, Johnson and Nyquist independently developed a quantitative framework: the mean-square voltage $\langle V^2\rangle$ across a resistor $R$ is given by the expression $4k_{\mathrm{B}}TR\Delta f$, where $k_\mathrm{B}$ represents Boltzmann's constant, $T$ is the absolute temperature, and $\Delta f$ denotes the frequency bandwidth~\cite{johnson1928thermal,nyquist1928thermal}. This relationship established an irreducible thermal noise floor, now known as Johnson-Nyquist noise, which imposes fundamental limits on signal detection in electronic communication~\cite{dorfel2012early}. 
    
    \paragraph{\textit{Entropy, computation, and the cost of erasure}}\mbox{}\\[1pt]
    The statistical-mechanical study of thermal fluctuations and the engineering quest to suppress electronic noise seemed like independent pursuits until World War II entwined them inextricably. Wartime demands for secure telephony, long-range radio, and radar accelerated research in signal transmission so that messages could travel reliably through noisy channels. In 1948, Shannon, drawing on the probabilistic foundations laid by Gibbs and Boltzmann, modeled a communication link as a thermodynamic system and introduced information entropy, $H=-\sum_ip_i\log_2 p_i$, where $p_i$ is the probability of occurrence of each message, a direct analog of thermodynamic entropy~\cite{shannon1948mathematical}. He showed that additive noise expands the ensemble of possible messages and derived the channel-capacity bound $C=B\log_2\left(1+S/N\right)$, which sets the maximum error-free rate for a bandwidth $B$ and signal-to-noise ratio $S/N$. Shannon's insights welded noise and information together, revealing that noise itself shapes the very essence of information. 
    
    The realization that entropy underlies both thermal noise and Shannon uncertainty points to a deeper union between information processing and the physical laws of nature. This connection proved instrumental in resolving Maxwell's famous paradox: a demon that sorts fast- and slow-moving molecules could, in principle, lower a system's entropy without expending energy, apparently violating the second law of thermodynamics. Szilard (1929) argued that the demon's act of measurement and memory must itself carry a thermodynamic cost~\cite{szilard1929entropieverminderung}. Landauer (1961) quantified that cost, showing that erasing a single bit of information irreversibly dissipates at least $k_{\mathrm{B}}T\ln 2$ joules of heat~\cite{landauer1961irreversibility}, a result that led him to the dictum \emph{information is physical}~\cite{landauer1991information}. Bennett and colleagues subsequently generalized Landauer's argument, demonstrating that only logically irreversible operations incur this energetic penalty and that, in principle, computation carried out by reversible gates can approach arbitrarily low dissipation, thereby tying the ultimate limits of computing and communication to the second law of thermodynamics~\cite{bennett1987demons}.
    
    \paragraph{\textit{Noise driven order out of equilibrium}}\mbox{}\\[1pt]
    If noise underpins information, can it also give rise to structure? In his pioneering work on dissipative structures~\cite{prigogine1955thermodynamics,prigogine1967symmetry,prigogine1978time}, Prigogine showed that a steady influx of energy can amplify microscopic fluctuations, selecting macroscopic modes that export entropy through heat or chemical flux, thus breaking symmetries where linear theory predicts only decay~\cite{kondepudi1981sensitivity,wood1985quantitative}. Echoing Landauer's principle, where erasing information incurs an entropic cost, such systems \emph{erase} less-stable configurations through dissipation, and the resulting entropy is carried away as heat.
    
    Two decades later, a complementary route to complexity appeared as noise became the very seed of self-organization. In 1987, Bak, Tang, and Wiesenfeld proposed self-organized criticality as a universal route to complexity~\cite{bak1987self}. In their sand-pile model, the slow, continual addition of grains drives the system to a critical state without external tuning, and the ensuing avalanches display power-law size distributions and generate scale-free $1/f$ (pink) noise. Similar $1/f$ spectra and the accompanying bursty dynamics appear in diverse phenomena ranging from neuronal avalanches~\cite{beggs2003neuronal} and earthquakes~\cite{bak2002unified} to stock-price fluctuations~\cite{black1986noise,gopikrishnan1999scaling}, reinforcing self-organized criticality as a hallmark of systems poised on the cusp of order and chaos.
    
    The ubiquity of $1/f$ spectra highlights that fluctuations are not peripheral but foundational to the dynamics of complex systems. Far from disrupting order, noise facilitates transitions across phase boundaries, help free systems from local energy traps, and coordinate interactions over long ranges. Therefore, it is no surprise that noise plays a pivotal role in the study of active matter.
    \end{minipage}
    \endgroup}
    \end{center}

\subsection*{Stochastic gene expression and cellular decision-making}
The central dogma of molecular biology, transcription of DNA into mRNA, followed by translation into proteins, operates at copy numbers so low that the arrival of a single mRNA can appreciably shift protein levels~\cite{elowitz2002stochastic,kaern2005stochasticity}. At such low molecule counts, even the presence of a single mRNA can lead to substantial cell-to-cell variability in protein levels~\cite{viney2013adaptive,coulon2014kinetic}. Early hints came from Novick and Weiner~\cite{novick1957enzyme}, but rigorous exploration required reliable single-cell assays. Pioneering work by Elowitz \textit{et al.} employed fluorescent reporters to distinguish intrinsic noise - arising from stochastic events at individual genes - from extrinsic noise, reflecting broader physiological variations~\cite{elowitz2002stochastic}. Thattai and van Oudenaarden showed analytically that intrinsic bursts alone can destabilize cellular states~\cite{Thattai2001-rx}, while Assaf \textit{et al.} later quantified how extrinsic fluctuations modulate phenotypic switching at the population level~\cite{assaf2013extrinsic}, highlighting the substantial impact of noise on cellular decision-making.

\paragraph{\textit{Noise-induced adaptability}} Considering the evolutionary significance of stochasticity, cells do not merely endure noise but actively harness it through adaptive strategies such as bet-hedging, an approach that distributes phenotypic states to maximize survival in fluctuating environments~\cite{Thattai2001-rx,kussell2005bacterial}. Kussell and Leibler's geometric-growth argument shows that probabilistic switching can optimize long-term fitness even at short-term cost~\cite{kussell2005phenotypic}, while Xue and Leibler demonstrated that the same diversification buffers demographic variability~\cite{xue2017bet}. Empirical evidence from Bigger's discovery of antibiotic persister cells~\cite{bigger1944treatment} and Balaban \textit{et al.}'s subsequent single-cell tracking of their stochastic entry into dormancy~\cite{balaban2019definitions,manuse2021bacterial} confirms that noise-driven bet-hedging confers resilience under stress.

Let $P\left(\mathbf{n},t\right)$ be the probability that the cell contains the copy-number vector $\mathbf{n}=\left(n_\mathrm{mRNA},n_\mathrm{prot},\ldots\right)$ at time $t$. For well-mixed reactions the chemical master equation is given by,
    \begin{equation}
    \label{eq:cme}
    \frac{dP\left(\mathbf{n},t\right)}{dt}=\sum_r\left[a_r\left(\mathbf{n}-\mathbf{\nu}_r\right)P\left(\mathbf{n}-\mathbf{\nu}_r,t\right)-a_r\left(\mathbf{n}\right)P\left(\mathbf{n},t\right)\right]
    \end{equation}
where $a_r$ is the propensity of reaction $r$ and $\mathbf{\nu}_r$ its stoichiometric jump~\cite{gillespie1977exact}. Exact Gillespie simulations reproduce Eqn.~\eqref{eq:cme} event-by-event without approximation. In the mesoscopic regime, each reaction fires many times over the integration step. A van Kampen system-size expansion smooths discrete jumps into diffusive Gaussian white noise, yielding the chemical Langevin equation,  
    \begin{equation}
    \label{eq:cle}
    \frac{d\mathbf{X}}{dt}
       =\sum_{r}\mathbf{\nu}_r a_r\left(\mathbf {X}\right)
       +\sum_{r}\mathbf{\nu}_r\sqrt{a_r\left(\mathbf{X}\right)}\,\mathbf{\xi}_r\left(t\right)
    \end{equation}
where $\mathbf{X}$ is the continuous copy-number vector with independent white noises~\cite{wilkinson2018stochastic} $\langle\xi_{r,i}\left(t\right)\xi_{s,j}\left(t^\prime\right)\rangle=\delta_{rs}\delta_{ij}\,\delta\left(t-t^\prime\right)$, implying independence across reactions $\left(r,s\right)$, species $\left(i,j\right)$ and time. Because the reaction propensity $a_r(\mathbf X)$ depends on molecular state under mass-action kinetics, the noise is multiplicative (state-dependent). For a linear birth-death gene the steady-state mRNA counts are Poisson with standard deviation $\sim\sqrt{\mu}$. Promoter bursting leads to over-dispersion: whereas a Poisson process gives a Fano factor $F=1$, the telegraph model yields $\mathrm{Var}\left(m\right)=\mu+\mu^2/r$, and thus $F=\left(1+\mu/r\right)>1$. In particular, at fixed burst frequency $r$, the variance grows super-linearly with the mean (and the standard deviation scales $\sim\mu$ for large $\mu$). Once these noisy gene products leak into the extracellular milieu, they become shared signals that couple cells together, turning intracellular randomness into population-level communication.

At the collective scale, quorum sensing lets bacterial populations turn individual noise into coordinated decisions. At low cell density, stochastic production and diffusion of auto-inducers create large copy-number fluctuations, producing heterogeneous gene-expression states~\cite{waters2005quorum,bassler2006bacterially}. As density rises, molecules accumulate and the noise is effectively averaged out once a threshold is crossed, synchronously triggering collective behaviors such as biofilm formation, virulence-factor secretion, or sporulation~\cite{miller2001quorum,camilli2006bacterial}. Hense \textit{et al.} showed that this noisy build-up prevents premature activation, ensuring the costly response is triggered only when the signal is reliably above background~\cite{hense2007does}. Quorum sensing thus follows a two-step recipe: fluctuations are first amplified as exploratory variation, then integrated into a robust, all-or-none collective behavior, a vivid example of evolutionary adaptation through controlled stochasticity, a theme that recurs across scales we examine below.

\paragraph{\textit{Noise-induced multistability}} Fluctuations can carve two or more stable expression states out of the same genome, endowing clonal cells with diverse phenotypes~\cite{dubnau2006bistability}. Kussell and Leibler argued that such bistability pays off when environments switch unpredictably, just as thermal kicks help a particle hop over an energy barrier~\cite{kussell2005phenotypic}. Empirical investigations in \textit{E. coli} confirm the coexistence of multiple metabolic states driven by noise within the lactose utilization operon~\cite{ozbudak2004multistability,santillan2007origin}, while yeast promoters stochastically pick lineage fates~\cite{acar2005enhancement,pomerening2008uncovering}. Likewise, multistable networks extend this idea to multiple potential lineages or functional states, as observed in stem-cell differentiation~\cite{macarthur2012nanog,huang2009non}. 

Projecting the chemical Langevin description onto a single slow coordinate and reorganizing the drift as a gradient yields a Fokker-Planck equation~\cite{gardiner2009stochastic}. The regulatory interaction pattern sets the shape of the effective potential landscape, which may contain multiple wells; Kramers' theory~\cite{hanggi1990reaction} then converts the barrier heights into switching rates, while well depths quantify state stability and hysteresis~\cite{munsky2012using}. This continuum picture connects single-molecule simulations to analytical physics, showing how cellular circuits filter or amplify fluctuations, dampening unwanted noise while preserving reliable decision making~\cite{altschuler2010cellular}.

With that mechanistic view in place, these cell-level switches raise a broader question: how do noise-shaped phenotypes feed into long-term evolutionary dynamics? Over the past decade, in-situ single-molecule fluorescence and super-resolution imaging have enabled direct visualization of an organism's genetic material down to specific genes or gene segments~\cite{larson2011real,sanchez2013regulation,senecal2014transcription,zoller2015structure,wang2016real,Sanchez2019-yj,rodriguez2020transcription,Zoller2022-jz}. These spatial transcriptomic approaches have, in turn, yielded unprecedented datasets capturing noise-induced state transitions, while stochastic thermodynamics has shed light on how replication, variation, and selection generate order far from equilibrium. 

For instance, Rao and Leibler recast reproduction–variation–selection dynamics as nonequilibrium thermodynamic processes, revealing evolutionary forces that transcend simple fitness landscapes~\cite{rao2022evolutionary}. In a similar vein, Ravasio \textit{et al.} demonstrated that life can leverage order through speed, favoring faster replication to assemble structures otherwise unattainable at equilibrium~\cite{ravasio2024minimal}. Tlusty and Libchaber likewise envision life as a cascade of self-assembling machines~\cite{tlusty2025life}, channeling noise into multiscale self-organization. In parallel, Nitzan \textit{et al.} showed that correlated fluctuations preserve spatial information in transcriptomic data, highlighting the crucial role of spatiotemporal noise in cellular decision-making~\cite{nitzan2019gene}.

Noise carries our story from single-mRNA sparks to quorum-sensing choruses, painting a landscape whose very distribution of phenotypic states is shaped by evolution. Yet every landscape needs a cartographer. Enter Maxwell's demon, armed with a ledger to record each fluctuation and a phosphate coin to test whether the signal exceeds thermal noise, biasing irreversible transitions and triggering directed state changes in receptor sensing, proofreading accuracy, and chemotactic steering~\cite{davies2019demon}.

\begin{figure}[t]
    \centering
    \includegraphics[width=0.7\linewidth]{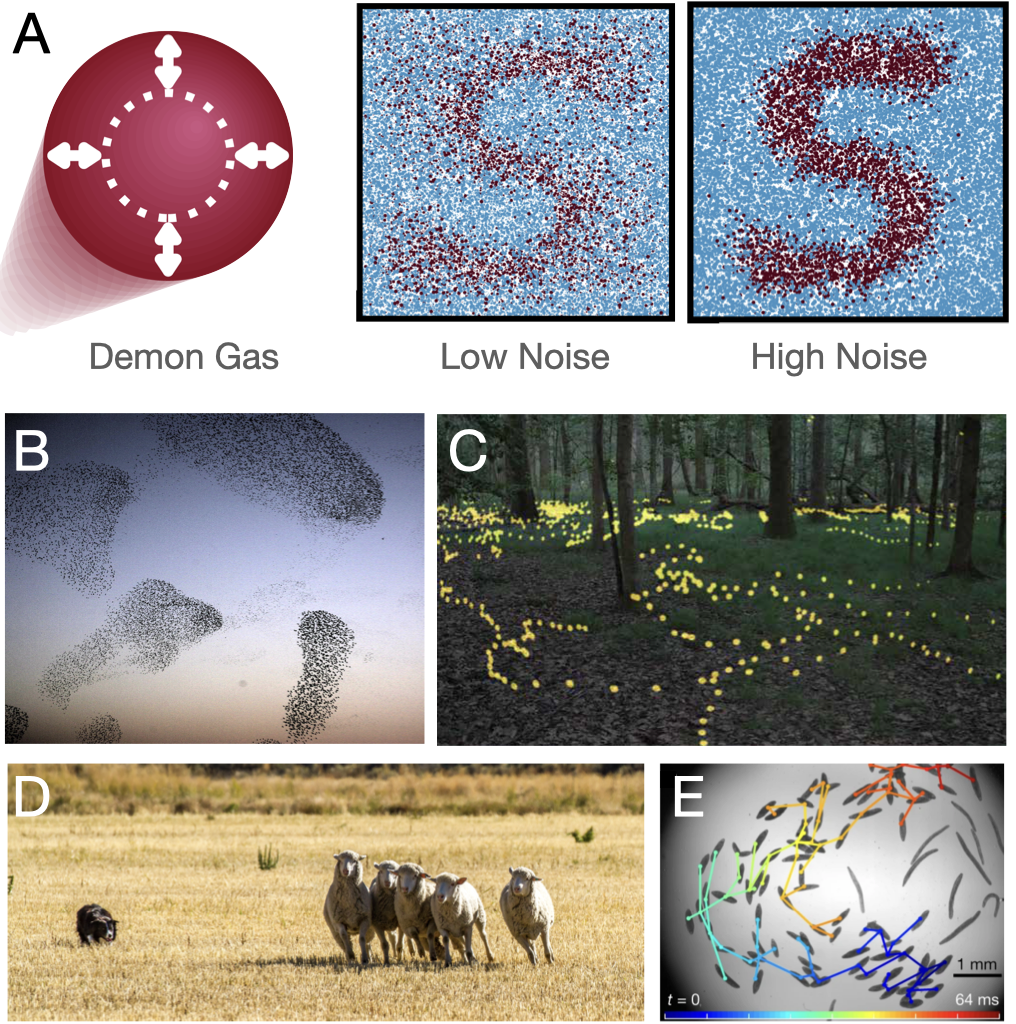}
    \caption{\textbf{Order from noise}. \textbf{A)} A demon gas generates patterns by harnessing thermal noise as an energy source~\cite{VanSaders2023-yl}; \textbf{B)} Massive starling flocks murmurate to evade predators, converting noise into large-scale coordinated motion~\cite{king2012murmurations}; \textbf{C)} Fireflies synchronize their flashes by using the fluctuating signals of nearby individuals as cues~\cite{peleg2018collective}; \textbf{D)} Small groups of sheep align in response to external noise, such as a scare from a shepherd dog~\cite{chakrabortty2025controlling}; \textbf{E)} Unicellular organisms like \textit{Spirostomum} use hydrodynamic fluctuations to generate synchronized contractions~\cite{mathijssen2019collective}.}
    \label{fig:noise_inf}
\end{figure}

\subsection*{Inference: demons that measure, remember, and decide} 
Biological signaling begins with an act of measurement. Whenever a receptor binds a ligand, the cell confronts the same question posed by Maxwell's mythical demon: \emph{Is this fluctuation informative enough to spend energy on?} The Berg-Purcell sensor answers by averaging stochastic binding events over time, writing the result into a transient methylation memory, and then hydrolyzing ATP to correct for the inevitable counting noise~\cite{berg1977physics,endres2009maximum,mehta2012energetic,kaizu2014berg,aquino2016know}. The energetic payment elevates accuracy beyond the equilibrium limit, formalizing an information-theoretic cost first quantified by Mora and Govern~\cite{mora2010limits,govern2014energy}.

The same sense-measure-act motif scales to the entire \textit{E. coli} chemotaxis pathway, where multiple demons coordinate their inferences~\cite{porter2011signal}. Ligand changes are encoded in receptor-methylation memory~\cite{vladimirov2009chemotaxis}; feedback between methylation and kinase activity filters environmental noise~\cite{hathcock2023nonequilibrium,Ito2015-ma}, and additional feedback modules bias CW to CCW flagellar switching~\cite{Tu2008-yl}; three nested demons thereby coordinate their inferences~\cite{porter2011signal}.

\paragraph{\textit{Error correction is inference in a different guise}} In addition to sensing, cells expend energy to proofread and correct errors. Hopfield–Ninio kinetic proofreading~\cite{hopfield1974kinetic,ninio1975kinetic} inserts an energy-driven discard branch that suppresses replication and translation errors at the cost of speed, embodying the accuracy-work trade-off quantified by Murugan \textit{et al.}~\cite{Murugan2012-ci,Murugan2014-lt}. In immune signaling, McKeithan applied the same scheme to T-cell receptors, and subsequent work by Feinerman \textit{et al.} quantified how the ATP-powered phosphorylation ladder raises the decision threshold, causing weak antigens to time-out before activation~\cite{mckeithan1995kinetic,feinerman2008quantitative}. By hydrolyzing one phosphate coin per proofreading step, these molecular demons continually retune speed, fidelity, and energy dissipation~\cite{cui2018identifying,boel2019omnipresent}.

Even with such proofreading, thermal agitation never disappears; yet molecular-recognition systems routinely squeeze astonishing fidelity from it, DNA replication errs~\cite{kunkel2004dna} only $\sim10^{-10}$ per base, translation~\cite{ninio1991connections,allan2009evolutionary} about $10^{-4}$, and chemotaxis can steer on differences of just a few ligand molecules. Crucially, the same circuitry can also turn noise into a strategic asset: transcription-factor affinities~\cite{Boeger2022-ax} differ by only a few $k_{\mathrm B}T$, promoters fire in stochastic bursts~\cite{leyes2023transcriptional}, and under stress those bursts widen, seeding phenotypic variants that survive antibiotics or starvation~\cite{Acar2008-ny,pal2024living}. In short, biomolecular circuits first pare errors to physical limits, then, when conditions demand, let fluctuations explore new states, with ATP expenditure tipping the balance between rigidity and adaptability.

\subsection*{Molecular motors: demons that pull against noise}
The demon that once was a bookkeeper now reveals its face. At nanometer scales inertia is negligible; if a motor pauses it coasts only angstroms, so equilibrium is physiological death. To stave off that fate, the demon binds ATP, tests a thermal fluctuation, and, only when the cargo is biased forward, snaps shut a molecular pawl~\cite{astumian1994fluctuation,julicher1997modeling}. Purcell's classic essay, \emph{Life at Low Reynolds Number} foretold this necessity: viscosity dominates, and movement requires an external energy source that breaks detailed balance~\cite{purcell1977}.

\paragraph{\textit{Mechanics and energetic cost}} In Maxwell's \textit{gedankenexperiment} the demon opens a trapdoor when a fast molecule arrives. Molecular motors enact the same logic on cytoskeletal tracks~\cite{Guo2014-ab,Murrell2015-ew}. Pioneering single-molecule work by Vale~\cite{vale1988formation,rice1999structural,vale2000way,vale2003molecular}, Howard~\cite{howard1997molecular,howard2002mechanics}, Svoboda and Block~\cite{svoboda1994force} showed that each kinesin or myosin head undergoes an ATP-triggered conformational change precisely when the elastic neck-linker is biased forward, rectifying Brownian motion step by step. Smoluchowski's and Feynman's ratchet arguments, updated by Tu~\cite{Tu2008-yl}, formalized why thermal kicks alone cannot yield net motion in a single heat bath; an energy input is mandatory~\cite{smoluchowski1927experimentell,feynman2015feynman}.

The first-law balance makes the thermodynamic costs explicit, $Fv+\dot{Q}_{\mathrm{diss}}=\dot{W}_{\mathrm{chem}}$. ATP hydrolysis supplies $\dot{W}_{\mathrm{chem}}$; if the chemical well runs dry the ledger closes, motion stops, and diffusion resumes. Even in their best regimes, motors convert only $40-60\%$ of that chemical power into useful work because part of the free-energy drop must be dissipated to enforce directionality~\cite{howard2002mechanics,schmiedl2008efficiency,hwang2018energetic}. Spectacular single-molecule data now resolve many details, yet key questions remain: how dynein's gear-like $\mathrm{AAA}^+$ ring coordinates~\cite{sakakibara2011molecular}, how load-dependent gating passes tension between the two kinesin heads~\cite{niitani2025kinetic}, and how strain propagates along the myosin lever arm~\cite{nie2014coupling}.

\paragraph{\textit{Models, open puzzles, and synthetic echoes}} Multi-state kinetic schemes translate every debit and credit in that ledger into observable velocity and run-length statistics. Optical trap experiments have mapped how distinct nucleotide states partition the kinesin step~\cite{block2007kinesin}; state-cycle models convert the same data into stall forces that match the predicted $6-7$ pN for an $8\,\mathrm{nm}$ step at $\Delta\mu\!\approx\!20\,k_{\mathrm B}T$ per ATP~\cite{kolomeisky2007molecular,kolomeisky2015motor,visscher1999single,carter2005mechanics}. A complementary coarse-grained view treats the motor as a flashing-ratchet Fokker–Planck system in which chemical switching periodically modulates an asymmetric potential while noise supplies the kicks; Prost \textit{et al.} recovered stall forces and load–velocity curves with that minimalist description~\cite{prost1994asymmetric}. Magnasco's two-potential model~\cite{magnasco1993forced} had already shown that a time-asymmetric landscape plus thermal noise suffices to break detailed balance, foreshadowing these mechanochemical cycles.

Once the information-to-energy exchange is recognized, the motor becomes a template. Chemists now build synthetic rotaxane shuttles that bias Brownian motion with chemical fuel, and theory cleanly separates the flows of energy and information driving the rotation~\cite{berna2005macroscopic,bruns2014rotaxane,amano2022insights}. Biological variants add further sophistication: allosteric sensors let kinesin, dynein and ATP-synthase read their own strain and gate forward steps accordingly~\cite{flatt2023abc}. Sometimes the external bath itself is active; colloidal Brownian motors immersed in a suspension of swimming bacteria can even harvest the bacteria's nonequilibrium noise to boost their own work output~\cite{paneru2022colossal}. From this perspective, each motor cycle is an information-thermodynamic transaction: measure displacement, erase disagreement, and pay in ATP to enforce the choice.

\begin{center}
    \textbf{Box 2 \textbar\ Particles to fields: continuum theories for active collectives}
    \fbox{
    \begingroup
    \footnotesize
    \begin{minipage}{0.95\textwidth}
    
    \paragraph{\textit{Vicsek model: the particle's-eye view}}\mbox{}\\[1pt]
    A minimal flock consists of $N$ point particles that self-propel at constant speed $v_0$ while aligning with noisy neighbors inside a metric radius $R$ (or a fixed topological set). At discrete time-steps $\Delta t$
    \begin{equation}
    \theta_i\left(t+\Delta t\right)=\arg\left(\sum_{j\in\mathcal N_i}\mathrm e^{\mathrm i\theta_j\left(t\right)}\right)
                         + \eta\,\xi_i\left(t\right)\qquad\text{and}\qquad
    \mathbf r_i\left(t+\Delta t\right)=\mathbf r_i\left(t\right)+v_0\Delta t\,
                            \left(\cos\theta_i,\sin\theta_i\right)
    \label{box2:eq_vicsek}
    \end{equation}
    where $\theta_i$ is the heading of particle $i$, $\xi_i$ is a uniformly distributed random angle in $\left(-\pi,\pi\right]$, and $\eta$ sets the noise amplitude. Boltzmann-style coarse-graining shows that the polarization order-parameter $m=\frac{1}{N}\left\langle\left|\sum_{i=1}^{N}\mathrm e^{\mathrm i\theta_i}\right|\right\rangle$, vanishes for $\eta>\eta_{\mathrm c}\left(\rho\right)$ and increases continuously as $\eta$ is lowered below this critical value, where the mean number density in $d$ spatial dimensions is $\rho=N/L^{d}$, for a system of linear size $L$. This marks a flocking transition accompanied by giant number fluctuations~\cite{vicsek1995novel}. A further continuum limit of the discrete dynamics yields the Toner-Tu hydrodynamic equations discussed below.
    
    \paragraph{\textit{Polar flocks (Toner-Tu)}}\mbox{}\\[1pt]
    For the coarse-grained velocity (polar order) field $\mathbf v\left(\mathbf r,t\right)$ one obtains:
    \begin{equation}
    \partial_t\mathbf v+\lambda_1\left(\mathbf v\cdot\nabla\right)\mathbf v
       =\alpha\,\mathbf v-\beta\left|\mathbf v\right|^{2}\mathbf v
       -\nabla P+D_v\nabla^{2}\mathbf v+\boldsymbol{\xi}
    \label{box2:eq_TT}
    \end{equation}
    where $\alpha\propto\left(\eta_\mathrm c-\eta\right)$ changes sign at the flocking transition, $\beta>0$ saturates the magnitude, and $D_v$ is an effective viscosity. Convection ($\lambda_1$) breaks the Mermin-Wagner bound, giving true long-range order in two dimensions and the experimentally observed $\sim n^{1/2}$ density fluctuations in starling flocks and midge swarms~\cite{toner1995long,cavagna2015flocking} (Fig.~\ref{fig:noise_inf}B).
    
    \paragraph{\textit{Active nematic bend instability}}\mbox{}\\[1pt]
    Replacing polar order by an apolar nematic tensor $Q\left(\mathbf r,t\right)$ and coupling to an incompressible Stokes flow $\mathbf u$ yields
    \begin{equation}
    \partial_t Q=-\Gamma\frac{\delta\mathcal F}{\delta Q}+\lambda E+\xi,\quad
    \eta\nabla^{2}\mathbf u-\nabla p=-\zeta\nabla Q,\quad
    \nabla\cdot\mathbf u=0
    \label{box2:eq_actnem}
    \end{equation}
    with strain-rate tensor $E$, Frank elasticity $K$, and active stress $\zeta$. Bend modes destabilize when $\left|\zeta\right|>\zeta_c\simeq4\pi^2K/L^2$, producing defect turbulence whose length scale is set by the active length $l_a\sim\sqrt{K/|\zeta|}$~\cite{thampi2014instabilities,shankar2019hydrodynamics}.
    
    \paragraph{\textit{Motility-induced phase separation (MIPS)}}\mbox{}\\[1pt]
    For isotropic self-propelled particles whose speed $v\left(\rho\right)$ decreases with density $\rho$, a two-field theory couples $\rho$ to a polarization $\mathbf P$:
    \begin{equation}
    \partial_t\rho=-\nabla\cdot\left(\rho\,v\left(\rho\right)\mathbf P\right)+D\left(\rho\right)\nabla^{2}\rho\qquad\text{and}\qquad
    \partial_t\mathbf P=-\frac{\mathbf P}{\tau}-\frac{v\left(\rho\right)}{2}\nabla\rho
                        +D_r\nabla^{2}\mathbf P+\boldsymbol{\eta}
    \label{box2:eq_mips}
    \end{equation}
    Linear stability gives a spinodal line when $\left(\partial v/\partial\rho\right)_{\rho_0}<-v\left(\rho_0\right)/\rho_0$, predicting the dense-droplet / dilute-gas coexistence seen in colloids, bacteria and vibrating coin-size robots~\cite{cates2015motility}.
    
    \paragraph{\textit{One hydrodynamic lens, many organisms}}\mbox{}\\[1pt]
    Equations share a backbone: a conserved density or momentum couples to an order parameter whose symmetry class: polar vector ($\mathbf v$), apolar scalar ($Q$) or isotropic scalar ($\rho$), dictates the instability. The activity parameter (alignment noise $\eta$, active stress $\zeta$, speed gradient $\partial v/\partial\rho$) injects microscopic energy; the leading restorative term (Landau damping, Frank elasticity, effective pressure) counteracts it. When drive exceeds damping the system crosses a universal threshold—flocking transition, bend instability, MIPS spinodal, and large correlation lengths, defect chaos or giant fluctuations emerge. Thus, the same hydrodynamic language captures \emph{molecular} active nematics in acto-myosin extracts, \emph{cellular} density waves in migrating epithelia, and \emph{organismal} criticality in bird flocks, revealing a continuous theoretical thread from nanometers to meters.
    \end{minipage}
    \endgroup}
    \end{center}

\subsection*{Demons in concert: cargo to condensates} 
The lone ledger-keeper now swells into a full orchestra: thousands of molecular demons strike the same note in unison. What begins as a $\mathrm{pN}$ flicker in a single kinesin or dynein head becomes network-scale stress once motors share cargo or filaments. 

\paragraph{\textit{Load-sharing demons}} Klumpp \textit{et al.} showed that several kinesins in parallel haul heavier loads and run longer than any lone motor~\cite{klumpp2005cooperative}; Reck-Peterson \textit{et al.} observed cargo that stays attached even when ATP or force fluctuates~\cite{reck2006single}. M\"uller and Mallik traced that robustness back to a stochastic tug-of-war: load sharing lets one demon hold tension while its neighbor re-primes~\cite{muller2008tug,mallik2013teamwork} (Fig.~\ref{fig:ant_motor_protein_ising}). Small fluctuations can even trigger occasional directional reversals, turning random load redistribution into an escape hatch that prevents lock-up~\cite{Guo2014-ab,Murrell2015-ew}. Each motor thus reads local force, spends a phosphate coin, and biases its next step, a distributed decision network of demons. Even more abstract models reach the same conclusion: Van Saders, Fruchart and Vitelli showed that agents which \emph{decide}, on the basis of noisy collisions, when to bias their next kick will self-flock with no alignment force at all, an informational ratchet that converts microscopic choices into macroscopic order~\cite{VanSaders2023-yl} (Fig.~\ref{fig:noise_inf}A).

Replace those point particles by flexible filaments and the bookkeeping scales again: Kinesin-5 and myosin-II cross-link and slide microtubules and actin, generating contractile stresses essential for spindles and morphogenesis~\cite{vale2003molecular,howard2002mechanics,gardel2008mechanical}. Counterintuitively, moderate stochastic stepping helps order emerge: models by J\"ulicher, Surrey and Kolomeisky showed that noise broadens the search of configuration space~\cite{julicher1997modeling,surrey2001physical,kolomeisky2007molecular}, and N\'ed\'elec \textit{et al.} confirmed in vitro that fluctuations drive asters, contractile rings and vortex swirls~\cite{nedelec2001dynamic,nedelec2002computer}. When motor-generated active stress $|\zeta|$ exceeds filament elasticity, networks cross an extensile-bend threshold predicted by Simha and Ramaswamy~\cite{aditi2002hydrodynamic} and visualized by S\'anchez~\cite{sanchez2012spontaneous}. Beyond that point, a sea of topological $+1/2$ and $-1/2$ defects erupts, yielding active turbulence with power-law spectra~\cite{thampi2013velocity,giomi2013defect}. Recent experiments with reconstituted actomyosin filaments reveal heavy-tailed stress-release events and $1/f$ noise, direct hallmarks of self-organized criticality, once network disorder and connectivity exceed a threshold~\cite{sun2025feedback}.

Characteristic speeds, defect densities, and energy spectra all collapse onto universal scalings (\textbf{Box 2}). How cells retune the motor ensemble via ATP supply, cross-linkers or filament turnover to jump between ordered arrays and turbulence remains an open problem relevant to spindle morphogenesis and cortical flows~\cite{Doostmohammadi2018-fb}. The very feedback that births turbulence can, in different geometries, freeze into clocks or dissolve into jams.

\paragraph{\textit{Phase-locking demons and crowd control}} Where demons are anchored to a cylindrical axoneme rather than a planar lattice, the outcome is not turbulence but a clock. A cilium packs $\sim10^4$ dynein demons; feedback between sliding and tension bootstraps their individual jitters into a self-sustained limit-cycle oscillator~\cite{lindemann2010flagellar,riedel2007molecular}. In viscous fluid, each beat transmits phase information as a long-range Blake tensor~\cite{blake1971note,brumley2012hydrodynamic}. Above a density threshold this coupling selects a metachronal wave that rectifies the otherwise reciprocal stroke and pumps fluid~\cite{niedermayer2008synchronization,guirao2007spontaneous,brumley2016long}. Experiments with colonial green alga (\emph{Volvox}) show that a dash of intrinsic jitter actually stabilizes the wave by avoiding frustrated locking~\cite{bruot2016realizing,pedley2016squirmers}.

Where hydrodynamic coupling weakens, the same demons manage steric traffic. Random arrival times and brief detours help cargo sidestep jams on shared tracks~\cite{nishinari2005intracellular,arpaug2020collective}. The principle scales to whole-cell layers: mild noise disperses traction stresses and suppresses jam-like arrest in epithelia~\cite{angelini2011glass,trepat2012cell}. Whether intracellular motor noise directly seeds these tissue-level jamming transitions or is merely one component of a multiscale feedback loop remains unresolved \cite{bi2015density,Doostmohammadi2018-fb}. 

When steric relief by motion is impossible, the demon changes strategy and reaches for phase separation. At yet larger scales, weak multivalent interactions and ATP-hungry helicases let hundreds of enzymatic demons build, fuse, or dissolve liquid-like condensates that relieve molecular crowding~\cite{brangwynne2015polymer,banani2017biomolecular}. Here, enzymatic turnover, rather than surface tension, sets droplet size. How stochastic enzymatic cycles couple to mesoscale phase behavior therefore stands as a frontier problem at the active/soft-matter interface. 

Still, noise alone cannot sculpt order; it needs a bookkeeper that prizes the \emph{right} correlations. From a single mRNA to a flock of filaments, every level of the living machine is patrolled by the same tireless accountant, trading \textit{negentropy} for information, one phosphate coin at a time. At the molecular scale, motors spend ATP to overcome extrinsic thermal noise; one tier up, organisms harness intrinsic decision noise to tip whole populations from one collective state to another. Bodies now sense through form and act through function, each movement inscribing fresh information into the neighbors' world.

\section*{Orchestrating noise into decisions}
Every living organism is a roaming sensor-actuator loop: it samples its world, stores incoming information in its transient memory, acts, and reacts. Nested in that loop sits a decision layer - an \emph{orchestrator} with an open ledger, that integrates noisy inputs, an organismal analog of Maxwell's demon. Here, the currency is no longer ATP but behavioral cues: a wing-beat, an antenna tap, a glint of bioluminescence, a droplet of pheromone - each a small act with a metabolic cost. When many such individuals interact, their actions braid thousands of noisy local transactions into one coherent, collective decision. Intrinsic variability is filtered at the individual scale, and extrinsic fluctuations are averaged across the group, allowing swarms to act with a precision no single body could achieve.

\subsection*{Noise-induced multistability}
Animal groups are constantly buffeted by background noise (extrinsic), such as gusts of wind, shifting light, and turbulent eddies, but the decisive nudges often come from within. Intrinsic noise wells up because no two agents sense or react in precisely the same way; limited visual acuity, variable pheromone thresholds, and idiosyncratic motivation can skew the collective wisdom of the group~\cite{lorenz2011social,boettiger2018noise}. Because this noise is state-dependent, it intensifies or wanes with group configuration, and can both seed and lock collective choices~\cite{horsthemke1984noise,jhawar2020noise2}. The orchestrator reads these tremors robe by robe and turns them into decisive action~\cite{ellis1953gregarious,matsumoto1983noise,deneubourg1995collective,bonabeau1999swarm,gu2025emergence}. 

\paragraph{\textit{The biblical marauder}} Few creatures dramatize intrinsic noise more vividly than desert-locust nymphs (\emph{Schistocerca gregaria}), legendary for their ability to strip fields bare in a single afternoon~\cite{world2020locust}. High-speed field movies reveal that a handful of nymphs wobble off course first; alignment feedback magnifies those hesitations until a march that was steady for hours can wheel within minutes. Here, the first stumble is nothing more than a stochastic wobble; intrinsic jitter supplies the seed that alignment feedback then magnifies into a column-wide turn.

Buhl \textit{et al.} quantified the effect and found that the flip rate peaks at intermediate crowd density: too sparse and the column fragments, too dense and it freezes~\cite{buhl2006disorder}. Yates \textit{et al.} captured the transition as a Fokker-Planck escape over a density-dependent barrier and showed that, paradoxically, a touch of extra randomness is required at low density to keep the swarm cohesive~\cite{yates2009united}. These sudden pivots therefore emerge from the swarm's own jitter rather than from any external gust or landmark.

\paragraph{\textit{A tale of two trails}} 
In a simple Y-maze, foraging ants must choose between two identical branches. Recruitment via pheromone is deterministic, abandonment is approximately Poissonian, and the only signal is the ants' own stochastic presence. Positive feedback then locks in whichever branch gains an initial majority~\cite{pasteels1987self,kirman1993ants,detrain2006self}, reducing the likelihood of reversal. Let $N$ be the total number of active foragers, $n$ the number on branch A ($N-n$ on branch B), $\alpha$ recruits per encounter, and $\beta$ spontaneous abandonment rate of a trail. Biancalani \textit{et al.} proposed to model the dynamics in the chemical-master language used for gene circuits~\cite{biancalani2014noise}:

\begin{equation}
    \frac{dP\left(n,t\right)}{dt} =
      \underbrace{\alpha\,\frac{\left(n-1\right)\left(N-n+1\right)}{N}\,P\left(n-1,t\right)}_{\text{recruit to A}}
    + \underbrace{\beta\,\left(n+1\right)\,P\left(n\!+\!1,t\right)}_{\text{abandon A}}
    - \left(\alpha\,\frac{n\left(N-n\right)}{N}+\beta\,n\right)\,P\left(n,t\right)
    \label{eqn:mul_noise}
\end{equation}

A van-Kampen expansion with $x=n/N$ turns Eqn.~\eqref{eqn:mul_noise} into a state-dependent Langevin, formally identical to the chemical Langevin used for transcriptional bursts (see Eqn.~\eqref{eq:cle}):

\begin{equation}
    \dot{x}= \alpha\,x\left(1-x\right)-\beta\,x
    \;+\;
    \sqrt{\frac{\alpha\,x\left(1-x\right)+\beta\,x}{N}}\;\xi\left(t\right),
    \quad\text{where}\quad
    \langle\xi\left(t\right)\xi\left(t^\prime\right)\rangle=\delta\left(t-t^\prime\right)
\end{equation}

The drift can create two wells, one per trail. The multiplicative noise amplitude $g\left(x\right)=\sqrt{\left(\alpha x\left(1-x\right)+\beta x\right)/N}$ decreases with group size as $N^{-1/2}$ overall, yet it increases when one branch dominates $\left(x\to1\right)$ because the abandonment term $\beta x$ inflates the fluctuations. For mesoscale collectives, large enough to form trails, yet small enough that $N^{-1/2}$ noise has not vanished, these state-dependent kicks are sufficient to nudge the system over the Kramers barrier, so the switching rate peaks at intermediate food influx, just as crowd density tunes the flip rate in locust swarms.

\paragraph{\textit{Noise below the surface}} Slip beneath a lagoon and the orchestrator tries to visually couple alignment. Schooling fish are a natural aquatic test-bed: they lack pheromones, rely on vision alone, and can switch from loose milling to polarized motion in seconds. In groups of cichlids (\emph{Etroplus suratensis}), Jhawar \textit{et al.} measured heading fluctuations and found that their variance grows as $1-m^{2}$, where $m$ is the polarization order parameter; the very act of aligning inflates the noise that may soon dissolve the formation~\cite{jhawar2020noise1}. Hidden-Markov and maximum entropy reconstructions in other species had hinted at such multiplicative kernels~\cite{katz2011inferring,herbert2011inferring,jiang2017identifying}; here, sequential one-by-one alignment lays them bare~\cite{jhawar2020noise2}. Thus, the same state-dependent noise that flips ant trails or locust swarms can, underwater, snap a milling cloud into a polarized swarm, and back again. Small visual jitter thus acts as a reversible toggle, echoing the density-tuned flips on land.

\paragraph{\textit{But quality of information matters}} A decade ago, Couzin and colleagues showed that schools of golden shiners (\emph{Notemigonus crysoleucas}) usually follow the numerical majority. Yet a small faction broadcasting a much stronger directional cue can briefly steer the entire shoal. That minority leverage disappears as soon as a pool of uninformed fish joins because their random motion dilutes the signal and restores majority rule~\cite{couzin2011uninformed}; hinting that a strong cue can momentarily outweigh head count.

Drone footage from the 2020 East African locust outbreak, paired with VR experiments that systematically dialed cue salience up or down, now strengthens that intuition (Fig.~\ref{fig:organismal_inf}B). Sayin \textit{et al.} found that desert-locust swarms ignore the average heading of nearby peers, an assumption baked into earlier, density-driven self-propelled-particle models of marching swarms~\cite{buhl2006disorder,yates2009united}. Instead, they track the single most salient visual cue, sometimes a distant landmark, sometimes a dense knot of peers, and sprint toward it~\cite{sayin2025behavioral}. When salience drops below a cognitive threshold, the barrier collapses, and the swarm flips. 

Order, therefore, emerges whenever coherent motion cues dominate noise, almost irrespective of density, and is captured by a ring-attractor neural model rather than a Vicsek-style alignment rule~\cite{vicsek1995novel}. Together, the fish and locust studies imply that cue credibility, not swarm size, decides when intrinsic noise will flip a group's choice. Testing how far this rule extends and tracing the neural circuits that weigh salience in other species remain open challenges.

\paragraph{\textit{Social ties set consensus}} Social hierarchy further reshapes the orchestrator's decision rules in eusocial collectives. In emigrating colonies of carpenter ants (\emph{Camponotus sanctus}), only mesoscale groups achieve cohesive emigration without fragmentation~\cite{rajendran2022ants}. While smaller colonies drown in randomness, larger ones form accidental quorums following a stubborn minority into a poorer nest once pheromone trails cross the threshold. Where collective payoffs are lower, such as for shelter-seeking cockroaches (\emph{Blattella germanica}), feedback weakens further and partial occupancy of several refuges becomes the norm~\cite{jeanson2005self}. Across deserts, lagoons and leaf-litter, the orchestrator balances state-dependent intrinsic noise, amplifies the first persuasive fluctuation through positive feedback, and relies on judicious information dilution, whether via na\"ive agents or low-salience cues, to keep the group from freezing onto a bad choice.  

\subsection*{Inference: measure, remember, and decide}
Donning the garb of an explorer, the orchestrator's task is no longer to choose between competing options, but to map an uncertain landscape in real time: Who sees danger first~\cite{treherne1981group}, where is pasture the richest~\cite{king2012selfish}, and how wide is the exit~\cite{helbing2000simulating}? The motif is familiar from molecular sensing and demons that measure, store, and decide. However, this time, the memory resides not in methyl groups; it is etched in fleeting alignments, rotating leaders, and trails of scent or sound.

\paragraph{\textit{Ripples on the lily-pond}} Confronted with a harsh, open pond where herons perch above, bass lurk below, and mates still need to be found, a loose assemblage of frogs has no cover; the only defense is to gauge the distance to the nearest neighbor, remember it for a heartbeat, and hop just far enough to avoid being the closest target. Hamilton's lily-pond thought experiment predicts the outcome: purely individualistic moves, amplified by positional noise, drive the frogs toward one another until a dense, roughly circular cluster materializes, order born purely from selfish interests and geometric constraints~\cite{hamilton1971geometry,foster2017safety,williams2018adaptation}. 

Once simple proximity rules have pulled the agents into a cluster, a richer playbook can unfold: add fleeting leaders and an external prod, and the herd learns to steer itself.

\paragraph{\textit{Woolen scouts}} In dense Merino sheep (\emph{Ovis aries}) flocks, the orchestrator alternates slow, dispersed grazing with brief, high-speed marches that dilute predation risk while sampling fresh pasture~\cite{king2012selfish}. During the slow phase, semi-random sub-groups wander in search of richer patches but also expose stragglers to predators~\cite{ginelli2015intermittent}. A march ignites when one sheep twitches; the cue propagates through a directed interaction network that rewires as hesitation grows, handing leadership to a new scout every few minutes~\cite{gomez2022intermittent}.

\begin{figure}[t]
    \centering
    \includegraphics[width=0.7\linewidth]{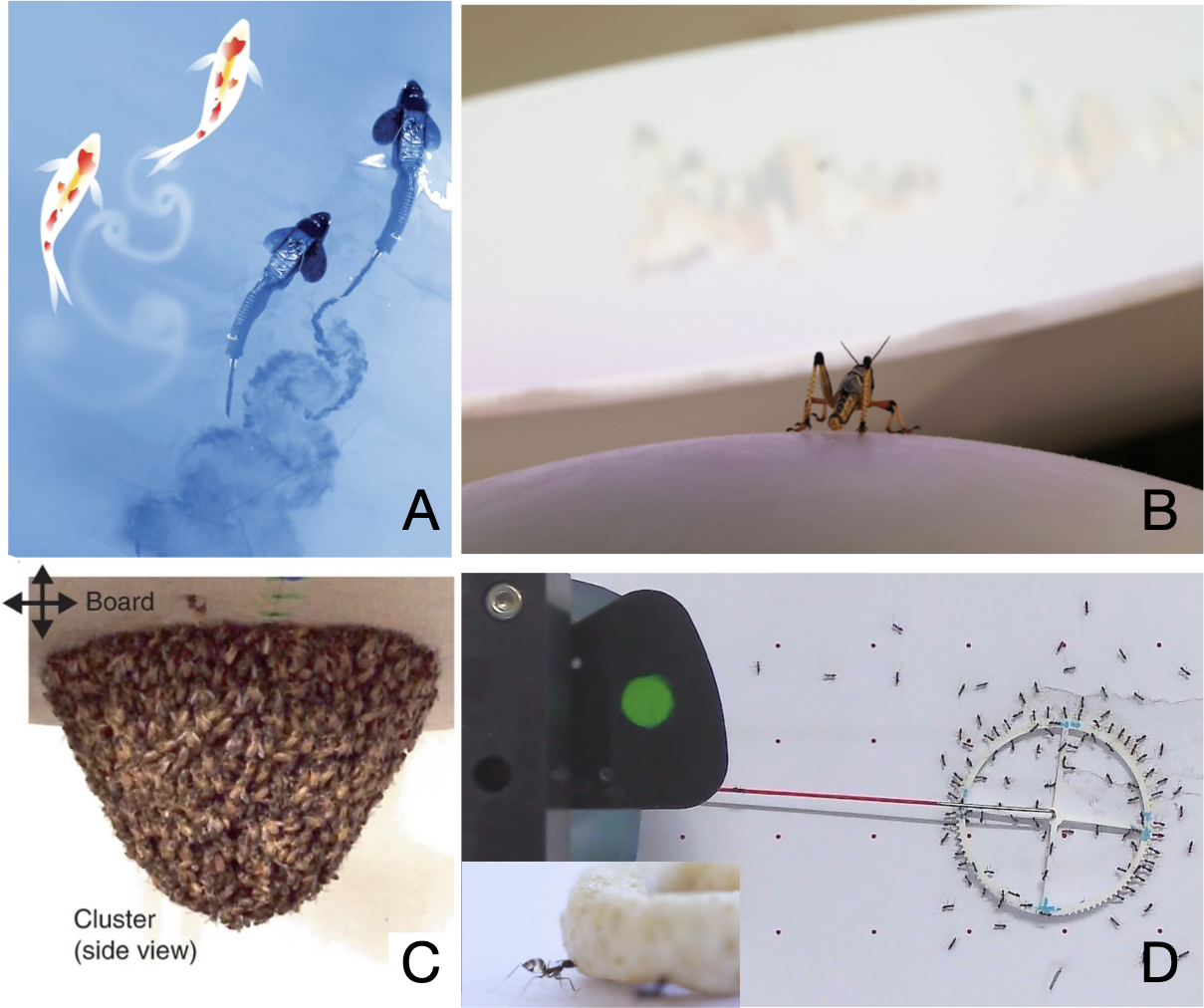}
    \caption{\textbf{Inferring behavior through mechanical and sensory inputs}. \textbf{A)} Real fish synchronize their swimming with a robotic fish to reduce energetic cost (picture credits: Couzin lab); \textbf{B)} A locust in a virtual reality arena learns to align its motion with projected neighbors (picture credits: Couzin lab); \textbf{C)} A honeybee swarm subjected to mechanical perturbations adapts its shape dynamically through structural morphogenesis~\cite{peleg2018collective}; \textbf{D)} A robot perturbs cooperative cargo transport in a group of longhorn ants, revealing a group-size-dependent collective susceptibility~\cite{chatterjee2025maximal}}
    \label{fig:organismal_inf}
\end{figure}

High-speed drone footage shows that a pursuing dog sharpens the self-compression into a front-to-back wave, pushing the flock toward a near-critical, high-synchrony state perfectly tuned for rapid escape~\cite{jadhav2024collective}. By contrast, in arena contests with just five sheep, Chakrabortty and Bhamla treat the dog as a control knob: a short dash followed by a pause, nudges the mesoscale flock into its own packing reflex without panic, demonstrating that stochastic nudge-and-wait inputs can steer noisy collectives with minimal probing~\cite{chakrabortty2025controlling} (Fig.~\ref{fig:noise_inf}D). Similarly, schooling fish employ a strategy known as vortex phase matching, adjusting timing of their tail beats to align with the flow structures created by nearby neighbors, thus conserving energy via hydrodynamic coupling~\cite{li2020vortex} (Fig.~\ref{fig:organismal_inf}A).

\paragraph{\textit{Memory in motion}} In longhorn ants (\emph{Paratrechina longicornis}) navigating a labyrinth, the orchestrator writes memory into pheromone trails. Lone scouts lay fleeting pheromone trails while hundreds of nest-mates prune weak paths and blaze strong ones, lighting the labyrinth like a living percolation graph.

\textit{Percolation theory tells us why this matters}: when the occupied-site fraction $p$ sits below the critical percolation threshold $p_c$, a random walker remains marooned on finite clusters (size $l$) as escape times grow super-diffusively, $\tau\sim l^\delta$ with fractal dimension $\delta>2$. Once $p$ nudges past $p_c$, a lattice-spanning backbone appears and traversal becomes near-ballistic, $\tau\sim l/v_{\mathrm{eff}}$, slashing journey time by orders of magnitude~\cite{nava1976electron,gefen1983anomalous,stanton1986analytic,de2009pg}. Gelblum \textit{et al.} showed that longhorn ant groups supply precisely this nudge through collective sensing while hauling cargo through a physical maze: a logarithmic bump in trail occupancy tips the colony into the ballistic regime, yielding an $\mathcal{O}\left(10^2\right)$ speed-up over passive random walks, contrasting de Gennes's \emph{ant in a labyrinth}~\cite{gelblum2020ant}. 

On soft substrates, the orchestrator emerges through simple digging rules reinforced by pheromone feedback, as in carpenter ants (\emph{C. pennsylvanicus}) or subterranean termites (\emph{Coptotermes gestroi})~\cite{deneubourg1995collective,deneubourg2002dynamics}. Scattered test bites nucleate a focused dig that breaches its enclosure the moment local erosion lifts the substrate itself above~\cite{prasath2022dynamics} $p_c$. Here, behavior and topology co-evolve through a stigmergic loop~\cite{theraulaz1999brief,dorigo2000ant}: simple digging rules, amplified by pheromone feedback, continually reshape the medium, and break the quenched-disorder assumption of classical percolation while turning noise into an excavation blueprint~\cite{tschinkel2004nest,buhl2005self}. 

Wrapped in stripes, the orchestrator aggregates noisy waggle runs as a honey-bee scout (\emph{Apis mellifera}). Each waggle run wobbles by several degrees, yet followers integrate dozens of runs, discarding outliers much like kinetic proof-reading, before flying the mean vector~\cite{tanner2008honey}. Finally, in honing pigeons (\emph{Columba livia}), na\"ive birds lead early exploratory loops; experienced birds assume command once a reliable map is etched, and leadership then oscillates to prevent premature lock-in on sub-optimal routes~\cite{valentini2021naive}.  

\paragraph{\textit{Brainless scouts}} Even without neurons, the orchestrator motif persists: measure, store, and decide via physical substrates. The slime mold \emph{Physarum polycephalum} thickens veins where nutrient pulses linger, solving mazes by flow-weighted memory~\cite{nakagaki2000maze}. Starfish larvae weave thousands of cilia into counter-rotating vortices; a flicker of shear re-orients the ciliary beat and funnels plankton to the mouth in milliseconds~\cite{gilpin2017vortex}. The placozoan (\emph{Trichoplax adhaerens}), a sheet of mono-ciliated cells behaves as a self-tuned active-elastic resonator: sub-second torque re-orientations let cilia flock, launch traveling traction waves, and steer the whole body toward food-agility built from mechanics, not synapses~\cite{bull2021ciliary,bull2021mobile}.

Rooted organisms show the same pattern in slow motion. Trading nerves for meristems, the orchestrator melts into a stand of sunflowers, rooted beings that still sense, remember, and choose~\cite{gruntman2017decision}. Noisy circumnutations, minute tip swings spanning three orders of magnitude in speed, act as functional noise; summed across neighbors, they steer the canopy from shade toward open sky~\cite{nguyen2024noisy}. Inside each cell, chloroplasts form an active glass: under dim light they jam, storing mechanical memory, yet a brighter flash fluidizes the layer and realigns the organelles within seconds~\cite{schramma2023chloroplasts}. Over days, wheat coleoptiles algebraically add and subtract overlapping gravity and touch cues on distinct timescales, and the costly elongation burst that follows shows that plants compute sums and differences, not mere integrals, of stimuli~\cite{riviere2023plants}. Even leaf-locked scouts, then, pool noisy signals across space and time, write them into mechanical or biochemical memories, and enact collective decisions that explore, defend, and forage.

Across croaking ponds, bleating fields, leafy canopies, pheromone labyrinths, and sun-streaked skies, the orchestrator measures through diverse sensors, remembers via ephemeral pheromone trails, visual alignments, waggle angles, or calcium pulses, and decides by averaging noisy samples into an estimate no lone explorer could muster. Yet the same logic guides a lonely traveler: a dung beetle (\emph{Scarabaeus satyrus}) rolling its prize beneath the Milky Way~\cite{dacke2013dung,foster2017stellar,khaldy2019effect}, or a desert ant (\emph{Cataglyphis}) tracing a zig-zag over shimmering salt~\cite{collett1998local,wehner2003desert}, must fuse the twinkle of celestial cues with the drift of their own stride count, distilling those noisy signals into a probabilistic compass that leads them home.

\subsection*{Noise-induced adaptability: living at the edge of criticality}
The orchestrator now moves among collectives that span meters. Where a few nearest-neighbor hops once sufficed to explore the entire decision space, that reality vanishes in these truly distributed swarms. No single agent sees the whole, so coordination must arise from long-range correlations that noise itself ferries through the crowd. By living at the edge of criticality, where correlation length rivals group size and susceptibility diverges~\cite{cipra1987introduction,landau2013statistical}, each noisy decision becomes a carrier wave, passing information from body to body and knitting the assembly into a single, responsive whole; the same physics appears even in solitary protists (\emph{Spirostomum ambiguum}), where ultra-fast hydrodynamic trigger waves race across millimeter-scale cells to coordinate contractions at rates that are hundreds of times faster than their basal swimming speed~\cite{mathijssen2019collective} (Fig.~\ref{fig:noise_inf}E).

\paragraph{\textit{Critical murmurs across air, water, and wax}} In European starling murmurations (\emph{Sturnus vulgaris}), each bird reacts to a fixed topological set of neighbors, letting heading changes propagate ballistically across hundreds of meters~\cite{ballerini2008interaction}. Such long-range order would violate the Mermin–Wagner theorem in equilibrium, yet self-propulsion and the full Toner-Tu quartet (\textbf{Box 2}), with a convective term $\lambda_{1}\left(\mathbf v\cdot\nabla\right)\mathbf v$, a pressure-like term $\nabla p$, a diffusive term $D\nabla^{2}\mathbf v$, and additive fluctuations $\boldsymbol{\xi}(t)$ override that limit and keep the flock supple rather than frozen~\cite{jenkins2022breaking,toner1995long,toner1998flocks}. High-speed, three-dimensional tracking confirms the prediction: velocity correlations are scale-free, so a single noisy twitch at one corner can race across the flock and let others pivot in near unison when a hawk dives~\cite{cavagna2010scale}.

    \begin{center}
    \textbf{Box 3 \textbar\ From magnets to collective choice}
    \fbox{
    \begingroup
    \footnotesize
    \begin{minipage}{0.95\textwidth}
    
    \paragraph{\textit{Binary choices across scales}}\mbox{}\\[1pt]
    Even the simplest cellular tasks hinge on binary decisions. A cargo vesicle hauled by opposing kinesin and dynein motors, for instance, flickers between plus-end and minus-end movement; each motor's choice to engage or disengage filters stochastic ATP turnovers into a directional vote that decides organelle positioning~\cite{welte2004bidirectional,ross2008cargo}. At organismal scales, \textit{P. longicornis} ants face a similar binary choice when collectively transporting food: pull (in the direction of local force) or lift (to reduce friction)~\cite{gelblum2015ant,feinerman2018physics}. Such binary decision making scenarios are ubiquitous, yet they somehow yield reliable, group-level consensus in noisy environments.

    \paragraph{\textit{Motor protein tug-of-war as an Ising-like switch}}\mbox{}\\[1pt]
    The tug-of-war can be coarse‑grained to a single coordinate, where stochastic attachment/detachment events govern the net number of active kinesins minus dyneins. Denote this difference by $n = N_k-N_d$. Under a mean-field load-sharing assumption, the instantaneous cargo velocity obeys  
    \begin{equation}
      \dot{x} \;=\; \frac{1}{\gamma}\left(N_kF_k-N_dF_d\right)\;+\;\sqrt{2D}\,\xi\left(t\right)
      \label{box3_eq:cargoLangevin}
    \end{equation}
    where $F_{k,d}$ are the single motor stall forces, $\gamma$ the viscous drag, $D$ an effective diffusion coefficient, and $\xi\left(t\right)$ Gaussian white noise. If we reorganize the load-dependent switch-on/off rates $k_{\pm}\left(n\right)$ as a gradient flow in an effective potential $U\left(n\right)$, Kramers' theory gives the run-length–determining switching rate  
    \begin{equation}
      k_{\text{switch}}
      \;=\;
      k_0\,\exp\!\left(-\,\Delta U\left(n\right)\big/k_{\mathrm B}T\right)
      \label{box3_eq:kramersMotor}
    \end{equation}
    with $\Delta U$ the energy barrier separating plus- and minus-end attractors. Together, Eqns.~\eqref{box3_eq:cargoLangevin} and~\eqref{box3_eq:kramersMotor} make explicit how molecular noise and motor copy number asymmetry set the fidelity of intracellular logistics~\cite{muller2008tug,kunwar2011mechanical}.

    \paragraph{\textit{Ising magnets to ant cooperative transport}}\mbox{}\\[1pt] 
    When \textit{P. longicornis} ants transport a Cheerio, each individual toggles between two states: pull ($p$) or lift ($l$) the cargo. Informed ants always pull toward the nest, acting like spins coupled to an external field; uninformed puller ants align with the momentary cargo direction, just as spins align with their neighbors, whereas lifters work to reduce friction by simply lifting the cargo at their end. The stochastic attachment/detachment dynamics are captured by microscopic spin-flips, and the rates of which are given by,
    \begin{equation}
      r_{p\leftrightarrow l}\;\propto\;\exp\!\left(\mp\,\mathbf{f}_{\mathrm{loc}}\!\cdot\!\widehat{\mathbf{p}}_i
      \big/ F_{\mathrm{ind}}\right),
    \end{equation}
    where $\mathbf{f}_{\mathrm{loc}}$ is the local force on ant $i$, $\widehat{\mathbf{p}}_i$ is its body-axis unit vector, and $F_{\mathrm{ind}}$ sets the effective coupling strength binding uninformed ants to the group~\cite{gelblum2015ant,feinerman2018physics}. As the number of informed ants (analogous to an external field) or the alignment strength $F_{\mathrm{ind}}$ crosses a threshold, the ensemble flips from disordered tug-of-war to ballistic coordinated motion, exactly the signature of an Ising-type phase-transition.

    \paragraph{\textit{Ising spin-lattice formalism}}\mbox{}\\[1pt] 
    The classical Ising model~\cite{landau2013statistical} treats each spin $s_i=\pm 1$ as a binary variable that tends to align with its neighbors while occasionally flipping due to thermal noise. The energy of an Ising lattice is given by
    \begin{equation}
      \mathcal{H}\;=\;-J\!\!\sum_{\langle i j\rangle}\! s_i s_j\;-\;h\!\!\sum_{i} s_i
      \quad\text{Glauber flip rate for $s_i$:}\quad w_{s_i\!\to\!-s_i} \;\propto\;\exp\!\left(-\beta\,\Delta E_i\right)\quad\text{where}\quad
      \Delta E_i = 2s_i\!\left(J\!\sum_{j\in\langle i\rangle}\!s_j + h\right)\quad\text{and}\quad\beta=1/k_\mathrm{B}T
    \end{equation}
    where $J>0$ promotes neighbor alignment and $h$ is an external field (e.g. an applied magnetic field or a bias toward one decision). When the reduced temperature satisfies $T/T_c<1$ (equivalently, $J/k_{\mathrm B}T > J_c/k_{\mathrm B}T_c$), the system undergoes a \emph{phase-transition}: a macroscopic magnetization $M=\langle s_i\rangle$ (the order parameter) emerges from microscopic alignment of the spins. Near $T_c$ the system is exquisitely sensitive, and small perturbations can tip it between ordered states, making the Ising model a natural metaphor for decision-making in noisy biological groups.

    \paragraph{\textit{Criticality amplifies consensus}}\mbox{}\\[1pt]
    At criticality, correlations span the entire lattice and fluctuations follow power-law statistics. In formal terms, the spin-spin correlation function
    \begin{equation}
      G(r)=\left\langle s_i s_{i+r}\right\rangle-\langle s_i\rangle^2\;\sim\; r^{-(d-2+\eta)}
    \end{equation}
    decays algebraically at $T_c$, so the correlation length $\lambda$ diverges ($\lambda\!\to\!\infty$). Here $r$ is the separation between spins, $d$ the spatial dimension, and $\eta$ the anomalous-dimension critical exponent. The same divergence appears in the field susceptibility
    \begin{equation}
        \chi \;=\;\frac{\partial M}{\partial h}\;=\; N\!\left(\langle M^2\rangle-\langle M\rangle^2\right),
    \end{equation}
    where $\chi$ measures the system's response to an external field $h$ which scales as $\chi\!\sim\!(T-T_c)^{-\gamma}$, where $\gamma$ is the susceptibility critical exponent. Biological analogues benefit from being poised at the cusp of criticality~\cite{romanczuk2023phase}. From neurons to bird flocks, living systems operate near critical points where correlation length and susceptibility are maximal. In cortical cultures and \textit{in vivo} recordings, neuronal avalanches exhibit power-law size and duration distributions, consistent with a branching process tuned to its critical point, thus maximizing dynamic range and information capacity~\cite{hengen2024criticality}. Starling murmurations and fish schools display scale-free velocity correlations and anomalously high susceptibility, hallmarks of a system perched at criticality~\cite{toner1995long,cavagna2010scale}. Some collectives may \emph{self-organize} to this state through slow driving and threshold-like release of stress or information, a mechanism known as self-organized criticality (SOC)~\cite{bak1987self,mora2011biological}. Whether achieved via explicit tuning or SOC, criticality endows biological groups with heightened sensitivity, broad dynamic range, and the ability to reach rapid, system-wide consensus in the face of noise.
    \end{minipage}
    \endgroup}
    \end{center}

Plunging into a living Gordian knot of California blackworms (\emph{Lumbriculus variegatus}), the orchestrator finds hundreds of supple bodies braided into a viscoelastic tangle, poised near the critical tangling index ($\approx 2$). A whisper of thermal or chemical noise unhooks only a few tails, yet those local link releases cascade through the blob, unleashing an ultra-fast topological trigger wave that lets it unknot and flow toward cooler, wetter ground, or burst apart to evade predators~\cite{patil2023ultrafast}. In sardine bait balls, a single startled fish can orchestrate an escape wave and split the school into fountain-like arcs that maximize escape speed without tearing social bonds~\cite{hall1986predator}. Drone footage and numerical models show that the split angle balances hydrodynamic thrust against social cohesion, as advection currents, pressure gradients, diffusion, and stochastic nudges again pin the school to a critical ridge~\cite{couzin2002collective,couzin2005effective,tunstrom2013collective,calovi2014swarming,bartashevich2024collective}. In honey bee clusters (\emph{Apis mellifera}), the orchestrator hangs beneath the swarm: only overloaded workers at the tip feel the extra weight, yet their grip adjustments percolate through the living cone and, when the swarm is perturbed, migrate the entire cluster up the strain gradient~\cite{sosna2019individual,peleg2018collective} (Fig.~\ref{fig:organismal_inf}C).

Across birds, midges, worms, fish, and bees, continuous tuning of spacing, speed, and grip keeps each group poised near the order-disorder threshold: scale-free correlations and finite-size scaling emerge~\cite{attanasi2014finite,hidalgo2014information,chate2014insect,romanczuk2023phase}, so a single agent's noisy decision can ripple through the entire collective without ever locking it into brittle order~\cite{mora2011biological,bialek2014social}.

\paragraph{\textit{The price of criticality}} But agility comes at a cost: extreme sensitivity invites false positives. Golden shiner schools (\emph{Notemigonus crysoleucas}) tuned too close to the critical threshold launch costly false-alarm escape waves even at harmless stimuli. When predators retreat, they slide back to a quieter, energy-saving sub-critical regime~\cite{poel2022subcritical}. Army-ant (\emph{Eciton}) living bridges solve the same dilemma by adding hysteresis: ants join a tense bridge more readily than they leave a slack one, filtering out high-frequency jitters, such as wind-shaken leaves or fleeting shadows, so the colony avoids fruitless rebuilding yet still responds to sustained changes in gap size~\cite{reid2015army,mccreery2022hysteresis}. In both cases, collectives either step back from the critical cusp or embed damping into their rules to temper the very sensitivity that usually empowers them.

\begin{figure}
    \centering
    \includegraphics[width=0.7\linewidth]{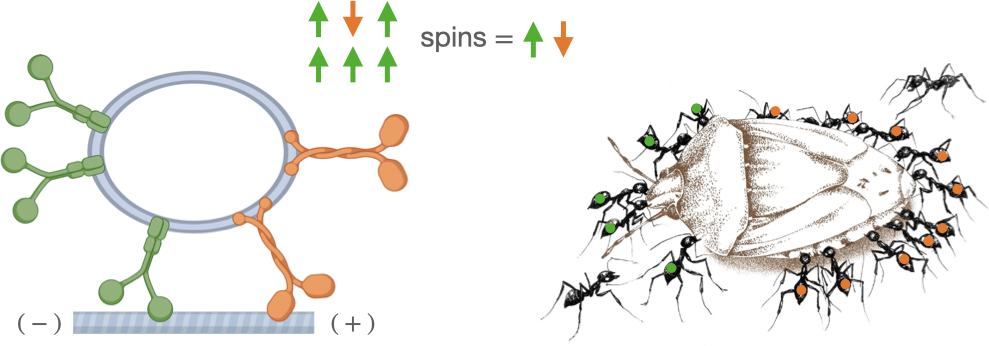}
    \caption{\textbf{Cooperative transport and the Ising model.} A molecular scale tug-of-war emerges when kinesin and dynein motors transport a vesicle along a microtubule (left). A similar tug-of-war arises at the organismal scale as longhorn ants cooperatively haul a dead insect in the wild (right). Ant illustration adapted from \textit{The Anteater's Culinary Guide} by Ofer Feinerman and Ehud Fonio, artwork by Sarit Bernard. The binary decision making dynamics in both can be modeled using the Ising model.}
    \label{fig:ant_motor_protein_ising}
\end{figure}

\paragraph{\textit{Tuning criticality in cooperative transport}} As the orchestrator clings to an oversized cargo, a new control knob-group size-sets the collective state. While small groups of longhorn ants (\emph{P. longicornis}) wander in a noisy random walk, larger ones surge ballistically toward the nest~\cite{gelblum2015ant,heckenthaler2023connecting}. Feinerman \textit{et al.} showed that each ant decides whether to pull or lift by sensing forces through the shared load (Fig.~\ref{fig:ant_motor_protein_ising}); an Ising-like model casts those microscopic choices as spin flips~\cite{gelblum2015ant,feinerman2018physics} (\textbf{Box 3}). These flips aggregate into mesoscopic force imbalances that transiently crown informed ants as leaders, steering the cargo home~\cite{gelblum2016emergent,ron2018bi}. 

Unlike the starling flocks, midge swarms, fish schools, and bee clusters, where criticality is self-organized, the longhorn ant system lets the orchestrator dial criticality by simply adding or removing workers. This external control turns cooperative transport into a living laboratory for testing the criticality hypothesis in biological collectives~\cite{romanczuk2023phase}. Chatterjee \textit{et al.} leveraged this tunable knob: disguised as a mechanical leader ant, the orchestrator injects a single bit of directional information through calibrated ant-scale forces (Fig.~\ref{fig:organismal_inf}D). Field experiments, supported by theory, reveal that collective susceptibility to such external perturbations peaks at an intermediate group size of roughly ten ants, large enough to average out the exploratory noise of smaller groups, yet small enough to avoid the inertia of larger ones~\cite{chatterjee2025maximal}.

\subsection*{Noise-induced adaptability: phase-locking across space and time}
The orchestrator now trades its explorer's map for a metronome. Success hinges not only on \emph{where} bodies sit but on \emph{when} they act. Local feedback, such as vision, sound, touch, and flow, retunes internal clocks much as metronomes phase-lock on a moving plank~\cite{kuramoto1984chemical,strogatz1993coupled,strogatz2001nonlinear,strogatz2005crowd,acebron2005kuramoto}. A minimal Kuramoto model with noise captures many such rhythms in nature:

\begin{equation}
    \frac{d\theta_i}{dt} = \omega_i + \frac{K}{N}\sum_{j=1}^{N}\sin\left(\theta_j-\theta_i\right) + \xi_i\left(t\right)
    \label{eq:kuramoto}
\end{equation}

\noindent
where $\theta_i$ is the phase of oscillator $i$, $\omega_i$ its intrinsic frequency, $K$ the coupling strength, and $\xi_i\left(t\right)$ a stochastic drive. Far from a nuisance, this noise is often the very spark that sets spatially distributed oscillators clicking into step. Weighting the sum by distance, line of sight, or acoustic direction reproduces the partial synchrony, chimeras, traveling waves, and danger relays that the orchestrator is about to witness.

\paragraph{\textit{Timing can spark}} The orchestrator lights the night with synchronous fireflies (\emph{Photinus carolinus}): flashes born deep inside the swarm ripple outward along clear lines of sight until thousands blink in unison~\cite{buck1938synchronous,sarfati2021self}; noise in who fires first is precisely what seeds the cascade (Fig.~\ref{fig:noise_inf}C). In \emph{P. frontalis}, the very same rules freeze the crowd into two interlaced sub-populations with a fixed phase lag, a living chimera that withstands wind, foliage, and moonlight~\cite{sarfati2022chimera}. Trading glow for chirping wings, the orchestrator tunes insect choruses: tree-crickets shorten a chirp when a neighbor leads and lengthen it when the neighbor lags~\cite{walker1969acoustic}; katydids of the genus \emph{Mecopoda} add period-doubling twists captured by a Poincar\'e map~\cite{sismondo1990synchronous}; and bush-crickets collectively stretch or squeeze their intervals to preserve the $12-18~\mathrm{dB}$ silence gap that choosy females demand, the heart of Greenfield's signal-preservation hypothesis~\cite{greenfield2008mechanisms,greenfield2017evolution,greenfield2023coordinated}. 

Zooming out from single choirs to whole soundscapes, dawn choruses reveal the same clock-tuning: songbirds (\emph{Zonotrichia albicollis}) burst into song within moments of civil twilight~\cite{staicer1996dawn}, periodical cicadas (\emph{Magicicada} spp.) ignite a minute-long acoustic avalanche at first light, an onset coordinated by noise and marked by a narrow susceptibility peak~\cite{goldstein2025photometric}; and howler-monkey (\emph{Alouatta palliata}) troops phase-lock their sunrise roars, overlapping barks that stitch distant groups into a single acoustic arena~\cite{de2015production}. In dense aerial traffic, the orchestrator manages acoustic jamming by echolocating bats (\emph{Pipistrellus kuhlii}); when calls collide, individuals boost intensity, duration, and repetition rate rather than merely shifting pitch to lift signal-to-noise in the melee~\cite{cvikel2015board,cvikel2015bats}.

\paragraph{\textit{Timing can defend}} On calm tropical seas the orchestrator launches alarm relays with water-striders (\emph{Halobates robustus}); leg-borne ripples outrun an approaching fish and buy every insect an extra split-second to scatter~\cite{treherne1982group}. The wave is triggered by a chance foot-tap; noise in who reacts first becomes the flotilla's alarm system. High above, the orchestrator recruits a shimmering wave in giant honeybees (\emph{Apis dorsata}); a stray abdomen flick can relay danger across an entire comb without a single bee taking flight~\cite{kastberger2012join}. Noise in who reacts first to a predatory wasp becomes the comb's broadcast channel. Similar shimmering surface waves in schooling sulphur mollies (\emph{Poecilia sulphuraria}) delay aerial attacks and reduce capture probability, supporting an anti‑predator role for wave‑based signaling~\cite{doran2022fish}.

The orchestrator also routes kinematic relays through vision: in schooling golden shiners (\emph{N. crysoleucas}), posture waves cascade through directed visual networks and re-orient hundreds in mere milliseconds~\cite{rosenthal2015revealing}; the first off-kilter tail beat is stochastic, and directed sightlines amplify it. Finally, the orchestrator locks phase underground with prime-number cicadas: soil-temperature cues act like a quenched field in an Ising-like consensus, and once the threshold is crossed, even moderate noise cannot break phase coherence, producing a synchronized emergence that overwhelms predators~\cite{hoppensteadt1976synchronization,sheppard2020self,goldstein2024cicadas}.

Across flashes, chirps, calls, ripples, waves, and even decade-long silences, shifts in phase, nudged by noise and guided by topological lines of sight, acoustic cones, or flow shadows, knit countless voices into a chorus that speaks in perfect time, a perfection sculpted by noise itself, not by its absence.

\section*{Outlook}
In 1877, Boltzmann taught us to read order in the statistics of disorder. However, Maxwell's mischievous demon soon unsettled that account by redefining information as a thermodynamic currency, one that Shannon and Landauer, almost a century later, priced at $k_\mathrm{B}T\ln2$ for each bit erased. Fast-forward to the dawn of the new millennium, and the pioneers of active matter physics showed that continuous, locally supplied energy lets real-world demons, from molecular motors and crawling cells to foraging ants and flocking birds, keep Boltzmann's order not merely alive but adaptive, amplifying the necessary fluctuations while discarding the rest. The next step is to balance the ledger across scales. 

In prebiotic autocatalytic networks~\cite{matreux2024heat}, stochastic fluctuations could have ratcheted free energy into sequence information, nudging protocells across the earliest kinetic barriers to heredity~\cite{kosc2025thermodynamic}. Applying modern stochastic thermodynamic tools to laboratory pre-RNA systems could soon price the very first phosphate coins life ever spent~\cite{gagrani2025evolution}, showing how the same ledger runs unbroken from the origin of life to today's molecular machinery. Leap ahead four billion years, and those same accounting principles guide single-molecule experiments~\cite{ashkin1986observation}. Calorimetric force spectroscopy resolves heat and work at single-base-pair resolution, while low-temperature assays expose glass-like RNA transitions~\cite{rissone2025dna,rissone2024universal}. High-throughput imaging of tens of thousands of DNA-unwinding events translates sequence diversity into an exchange rate between free energy and entropy dissipation~\cite{aguirre2024massively}, while a fuzzy sequencing-by-synthesis platform more than doubles the information efficiency of conventional cyclic terminator chemistry, turning each nucleotide addition into a higher-yield entropy-information transaction~\cite{zhou2025fuzzy}. Taken together, these tools complete the ledger at the nucleotide level: heat in, information out, setting the stage for the next rung of the hierarchy. 

\begin{center}
    \textbf{Box 4 \textbar\ Exorcising the demon}
    \fbox{
    \begingroup\footnotesize
    \begin{minipage}{0.95\textwidth}
    \paragraph{\textit{How does one hunt a Maxwell's demon?}}\mbox{}\\[1pt]
    Szilard's 1929 analysis revealed that the demon's act of measurement carries an entropy price~\cite{szilard1929entropieverminderung}; Landauer later showed that erasing the demon's memory costs at least $k_\mathrm BT\ln 2$ of heat~\cite{landauer1961irreversibility}. Hence any putative demon betrays itself through measurable entropy production. Consider a Markov system with micro-states $s_1, s_2... s_n$, where transitions from $s_i$ to $s_j\,\left(i\neq j\right)$ occur with propensities $w_{ij}\equiv w\left(s_i\!\to s_j\right)$. The state probabilities $P\left(s_i,t\right)$ then satisfy the following master equation

    \begin{equation}
      \frac{dP\left(s_i,t\right)}{dt}
      =\sum_{j\neq i}\left[w\left(s_j\!\rightarrow\! s_i\right)P\left(s_j,t\right)-w\left(s_i\!\rightarrow\! s_j\right)P\left(s_i,t\right)\right]
      \label{box4:eq1}
    \end{equation}
    
    Each jump from $s_i$ to $s_j$ produces a measurable entropy
    
    \begin{equation}
      \Delta\sigma_{ij}
      = k_{\mathrm B}\ln\!\frac{w\left(s_j\!\rightarrow\! s_i\right)P\left(s_j,t\right)}{w\left(s_i\!\rightarrow\! s_j\right)P\left(s_i,t\right)}
      \qquad\text{(a positive $\Delta\sigma_{ij}$ corresponds to net heat dissipation to the environment)}
      \label{box4:eq2}
    \end{equation}
    
    where $k_{\mathrm B}$ is the Boltzmann constant. Summing $\Delta\sigma_{ij}$ along a trajectory yields the total path-wise entropy production, the observable fingerprint of any Maxwell-like feedback or information-processing agent hidden in the dynamics. In practice, most microstates and many transition paths remain hidden. The tools below infer entropy production from partial information, each with its own assumptions and limitations.

    \setlength{\tabcolsep}{4pt}
    \renewcommand{\arraystretch}{1.15}
    
    \centering
    \begin{tabularx}{\linewidth}{@{}>{\bfseries}p{0.24\linewidth} 
                             >{\raggedright\arraybackslash}X 
                             >{\raggedright\arraybackslash}p{0.14\linewidth}
                             >{\raggedright\arraybackslash}p{0.12\linewidth}
                             >{\raggedright\arraybackslash}p{0.15\linewidth}@{}}
    \toprule
    \multicolumn{1}{@{}l}{\textbf{Thermodynamic tool}} &
    \textbf{Key relation} &
    \textbf{Regime} &
    \textbf{Dynamics} &
    \textbf{Data required} \\ \midrule
    \textbf{Fluctuation Dissipation Theorem (FDT)~\cite{kubo1966fluctuation}} &
    $\widetilde C(\omega)=\dfrac{2k_{\mathrm B}T}{\omega}\,\widetilde\chi^\prime{^\prime}(\omega)$ &
    Equilibrium \& near-equilibrium &
    Detailed balance &
    Linear response \\

    \textbf{Jarzynski \& related fluctuation theorems (Fig.~\ref{fig:microscale_inf}A)~\cite{jarzynski1997nonequilibrium,crooks1998nonequilibrium,hatano2001steady}} &
    $\displaystyle\left\langle e^{-\beta W}\right\rangle=e^{-\beta\Delta F}$ &
    Equilibrium initial distribution &
    Arbitrary time-dependent protocol &
    Work distribution $P\left(W\right)$ \\
    
    \textbf{Thermodynamic Uncertainty Relation (TUR)~\cite{barato2015thermodynamic,horowitz2020thermodynamic}} &
    $\frac{\mathrm{Var}(J)}{\langle J\rangle^{2}}
          \ge\frac{2k_{\mathrm B}}{\langle\sigma\rangle}$ &
    Non Equilibrium Steady State (NESS) &
    None (Markov) &
    Trajectory-level current $J$ \\
    
    \textbf{Cycle-current entropy production rate~\cite{skinner2021estimating}} &
    $\dot\sigma=k_{\mathrm B}\!\sum_{i<j}
            \left(\phi_{ij}-\phi_{ji}\right)\ln\!\frac{\phi_{ij}}{\phi_{ji}}$ &
    NESS &
    Markov &
    Probability currents $\phi_{ij}$ \\
    
    \textbf{Variance Sum Rule (VSR)~\cite{Di-Terlizzi2024-pr}} &
    $V_{\!\Delta x}\left(t\right)+\mu^{2}V_{\!\Sigma_F}\left(t\right)
       =2Dt+S\left(t\right)$\quad
    $S\left(t\right)=2\mu\!\int_{0}^{t}\!\left(C_{xF}-C_{Fx}\right)\,\mathrm dt^\prime$ &
    NESS &
    Langevin &
    Variances $V$, covariances $C$ \\ \bottomrule
    \end{tabularx}
    \end{minipage}
    \endgroup
    }
    \end{center}

\begin{figure}[t]
    \centering
    \includegraphics[width=0.7\linewidth]{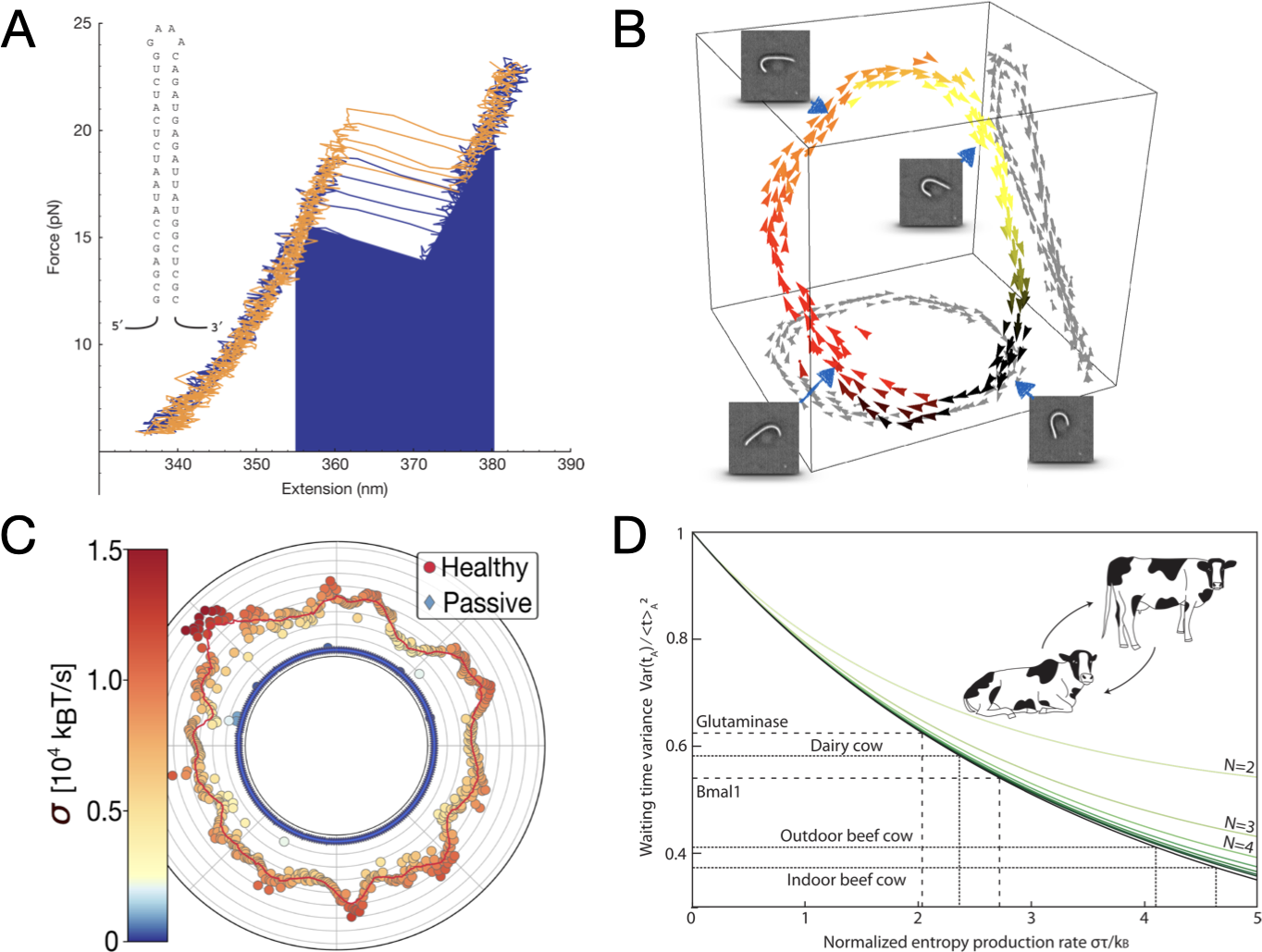}
    \caption{\textbf{Molecular inference through thermal noise and energy dissipation}. \textbf{A)} Force-extension curves of an RNA hairpin reveal the stochasticity of unfolding and refolding trajectories, with work dissipated or recovered (blue area) as the molecule transitions between states~\cite{collin2005verification}; \textbf{B)} Flagellar beating in \textit{Chlamydomonas} exemplifies a nonequilibrium steady state that violates detailed balance, producing cyclic probability flux loops in a low-dimensional phase-space~\cite{gnesotto2018broken}; \textbf{C)} Entropy production profiles of healthy and passive human red blood cells measured using optical tweezers and optical microscopy~\cite{Di-Terlizzi2024-pr}; \textbf{D)} Entropy production rate inferred from waiting time variance in a dissipative system: cows lying down before standing~\cite{skinner2021estimating}.}
    \label{fig:microscale_inf}
\end{figure}

What, then, is the energetic bill for harnessing noise? Three experimental lenses have begun to translate those phosphate coins into measurable joules. Tracking nanoscale shear in actomyosin gels, Mizuno \textit{et al.} found an excess of low-frequency motion that matches the heat a myosin motor must dump per ATP hydrolyzed~\cite{mizuno2007nonequilibrium}. Seara \textit{et al.} pushed the same idea to contracting gels, mapping where dissipation peaks as the network stiffens~\cite{seara2018entropy}. At the organelle scale, Battle, Turlier, and colleagues imaged beating flagella~\cite{Battle2016-ft} and flickering red blood cell membranes~\cite{turlier2016equilibrium}, then reconstructed the probability currents that swirl through phase space, directly visualizing the ledger's irreversible spend (Fig.~\ref{fig:microscale_inf}B). When such currents hide, Terlizzi \textit{et al.} showed that a simple variance asymmetry can still expose hidden entropy production by comparing how often trajectories run forward versus backward in time~\cite{Di-Terlizzi2024-pr} (Fig.~\ref{fig:microscale_inf}C). 

Because direct current mapping is invasive and data-hungry~\cite{martin2001comparison,Lynn2021-yp}, newer inference methods extract dissipation from partial information: coarse-grained state reconstruction~\cite{bilotto2021excess,dieball2022mathematical}, waiting-time statistics for non-Markovian jumps~\cite{skinner2021estimating,van2022thermodynamic,harunari2022learn}, and thermodynamic-uncertainty relations (TURs) that translate fluctuation levels into lower bounds on dissipation~\cite{kawai2007dissipation,nitzan2023universal} (Fig.~\ref{fig:microscale_inf}D). The Barato-Seifert bound, refined by Gingrich, Horowitz and co-workers~\cite{barato2015thermodynamic,horowitz2020thermodynamic}, makes the bookkeeping concrete: halving relative fluctuations must at least double entropy production, while irreducible timing noise sets the ledger's final precision floor (\textbf{Box 4}). 

Having priced dissipation at the single-cell level, we can now ask how the same fluctuations scale up to assemble whole organisms. Theory and imaging suggest that quirks in cell size, adhesion, and division timing can nudge unicells into nascent snowflake clusters. Under nutrient limitation, a single \emph{ACE2} loss-of-function in yeast (\textit{S. cerevisiae}) locks mothers to daughters, pushing clonally developing aggregates beyond the critical size where shared metabolites and stresses enter the ledger~\cite{ratcliff2012experimental,ratcliff2015origins,jacobeen2018cellular}. Once there, metabolism creates density gradients that drive self-generated flows, ferrying nutrients faster than diffusion and sustaining exponential growth; within a few hundred generations the clusters even evolve apoptotic division of labor, letting selection audit the ledger at the level of the collective and turning the molecular ledger into a tissue-level accountant on the road to true multicellularity~\cite{narayanasamy2025metabolically}. 

Zooming out even further to host-pathogen ecosystems~\cite{baym2016spatiotemporal}, stochastic thermodynamic models show that noise can accelerate the step-wise rise of antimicrobial resistance (AMR), whereas coupling between microbial communities can slow, or even forestall, it~\cite{hu2025effect} underscoring AMR as perhaps the defining thermodynamic race of our century~\cite{krishnaprasad2024antimicrobial}. The challenge ahead is to extend such trajectory-level thermodynamic inference from organelles to whole cells and ultimately, to tissues and even populations, so that we can directly price the energetic costs of sensing, locomotion, and decision making in living matter.

Extending this framework is more than just taxonomy. Decision making at every tier of biological organization draws from the same finite budget of free energy and information, yet spends that budget in strikingly different ways. While a cell shunts ATP to bias a signaling cascade, a colony reallocates thousands of bodies to bias a collective vote. Superorganisms and gregarious collectives thus act as living analogues of neural tissues~\cite{eckmann2007physics}, cognition emerging from local interactions among many noisy agents, providing an experimentally tractable mirror for the open questions of collective inference and consciousness that still elude us in the human brain. These systems therefore let us test a long-standing conjecture: adding layers of cognitive complexity does not linearly increase cooperation~\cite{dreyer2025comparing}, and can even dissipate information when individual incentives diverge~\cite{baltiansky2021dual}. Growing evidence now suggests that both cortical networks and insect swarms evade that pitfall by tuning themselves to the edge of criticality, where information travels afar while energetic costs stay bounded~\cite{romanczuk2023phase,hengen2024criticality,kemp2024information,muller2025critical,liu2025cognitive,chatterjee2025maximal}.
    
But studying full-scale organismal swarms in situ is logistically fraught~\cite{samson2020collective,hueschen2023wildebeest}. Tabletop robophysical swarms therefore promise a clean bridge between biological messiness and first-principles theory. Frugal Kilobots, particle robots, and modular tumbling cubes replay the same local rules that govern ants and termites or fish and cells, yet let us dial noise, connectivity, and feedback at will~\cite{rubenstein2012kilobot,rubenstein2014programmable,romanishin20153d,li2019particle}. Within this sandbox, discoveries predicted by active matter theory, such as first-order flocking, tricritical points, motility-induced phase separation, tunable non-reciprocal interactions, and low-rattling organization emerge as measurable phase diagrams rather than fleeting field anecdotes~\cite{giomi2013swarming,slavkov2018morphogenesis,savoie2019robot,fruchart2021non,chvykov2021low,wang2021emergent,ben2023morphological,saintyves2024self,newbolt2024flow}. Because artificial agents can be cloned by the hundred and rewired on the fly~\cite{yang2018grand}, robophysical collectives turn randomness from a nuisance into a programmable resource, feeding back fresh, testable hypotheses to biology.

The 2024 Nobel Prize~\cite{hopfield2024nobel} to Hopfield and Hinton reminds us that, whether in Boltzmann machines or beating hearts, intelligence blooms when energy landscapes let noise buy bits of insight. Framing active matter questions in the common currency of entropy production and information flow will therefore allow us chart a quantitative cost-benefit curve for cognition, from molecular switches to organismal-scale decisions, and bring the physics of life a step closer to the physics of mind~\cite{hidalgo2024eps,gompper20252025}.

\end{document}